\documentclass{iau}
\usepackage{graphicx}

\title[Toward Imaging Sub-Jovian Planets with \emph{JWST}] 
{Toward Direct Imaging of Low-Mass Gas-Giant Planets with the \emph{James Webb Space Telescope}}

\author[J.E. Schlieder et al.]   
{J. E. Schlieder$^{1}$, C. A. Beichman$^2$, M. R. Meyer$^3$, \& T. Greene$^4$}

\affiliation{$^1$NASA Postdoctoral Program Fellow, NASA Ames Research Center\\email: {\tt joshua.e.schlieder@nasa.gov}\\
$^2$NASA Exoplanet Science Institute/Caltech, $^3$ETH Z\"urich, $^4$NASA Ames Research Center }

\pubyear{2015}
\volume{314}  
\pagerange{119--126}
\jname{Young Stars \& Planets Near the Sun}
\editors{J. H. Kastner, B. Stelzer, \& S. A. Metchev, eds.}
\begin{document}

\maketitle

\begin{abstract}
In preparation for observations with the \emph{James Webb Space Telescope} (\emph{JWST}), we have identified 
new members of the nearby, young M dwarf sample and compiled an up to date list of these stars. Here we summarize 
our efforts to identify young M dwarfs, describe the current sample, and detail its demographics in the context of direct planet 
imaging. We also describe our investigations of the unprecedented sensitivity of the \emph{JWST} when imaging nearby, 
young M dwarfs. The \emph{JWST} is the only near term facility capable of routinely pushing direct imaging capabilities 
around M dwarfs to sub-Jovian masses and will provide key insight into questions regarding low-mass gas-giant properties, 
frequency, formation, and architectures.


\end{abstract}

\firstsection 
\section{Introduction}

M dwarfs, low-mass stars ($\le$0.6 M$_{\odot}$) cooler than the Sun ($\le$3900 K), are the most common stars in the Galaxy
($\sim$75\%). The youngest M dwarfs in the Solar neighborhood (d$\le$100 pc) are 
members of loose associations having common Galactic kinematics and ages $\sim$10-130 Myr; young moving groups (YMGs,
\cite[Zuckerman \& Song 2004]{Zuckerman2004}). Due to their proximity, youth, and intrinsically low-luminosities, 
M dwarf YMG members are prime targets for direct exoplanet imaging. Additionally, a sizable population of $\sim$Jupiter to $\sim$Neptune mass planets is detected at moderate separations ($\sim$1 - 10 AU) around M dwarfs by gravitational micro-lensing surveys (\cite[Cassan et al.~2012]{Cassan2012}).  For these reasons, many dedicated 
searches to identify the missing M dwarf members of YMGs have been undertaken, e.g. 
\cite[Kraus et al.~(2014)]{Kraus2014}, and several 
surveys to directly image their planets are ongoing (\cite[Bowler et al.~2015]{Bowler2015}). Despite the identification 
of several hundred likely new M dwarf YMG members and high-contrast imaging observations of a sizable 
fraction of that sample, YMGs still lack the large numbers of M dwarfs expected
(\cite[Kraus et al.~2014]{Kraus2014}) and only a few planetary-mass companions on wide orbits have been imaged. To probe the known population of low-mass gas-giants around M dwarfs via direct
imaging, more members of the nearby young sample must be identified and facilities with greater sensitivity are required. 

\section{A Search For Nearby Young M Dwarfs and the Current Sample}

We have identified nearly 400 candidate M dwarf YMG members using a selection algorithm that hinges on proper motion, 
photometry, and activity (see \cite[Schlieder et al.~2012]{Schlieder2012}). For the last 3 years we have pursued an all-sky,
spectroscopic, follow-up program to identify true YMG members in the sample that we call
\emph{CASTOFFS} (\cite[Schlieder et al.~2015]{Schlieder2015}). We have observed $\sim$$\frac{3}{4}$ of our candidates using 
high-resolution optical (MPG 2.2m/FEROS, CAHA 2.2m/CAFE) or medium-resolution near-IR (IRTF 3.0m/SpeX) 
spectroscopy and our analyses continue. Our high-resolution optical spectra have so far revealed more than 50 new 
M dwarf YMG members, several isolated young field M dwarfs, and a dozen spectroscopic binaries. Our analyses of the near-IR
spectra are just beginning and young M dwarfs with low surface-gravity features are already apparent 
in the data. 

We have compiled an up to date list of young, low-mass stars within 100 pc from the literature and added to 
it our new identifications from \emph{CASTOFFS}. We primarily include stars having both spectroscopic 
confirmation of youth and at least partial kinematics (proper motion and RV) consistent with YMG membership. A few well
characterized, young field stars are also included. From more than 30 literature references over the last $\sim$15 years and 
our new results, we find 440 $\sim$K5 - M9 systems (338 known or presumed single, 102 multiples) with ages $\sim$10-400 Myr at 
distances $\lesssim$100 pc. This is nearly a factor of five increase over the M dwarf YMG sample from \cite{Torres2008}. Only $\sim$$\frac{1}{3}$ of the sample has measured parallaxes, the remaining distances are kinematic or photometric 
estimates. There are fewer stars within 25 pc than anticipated and the population falls off very quickly beyond $\sim$M4 type and distances $\gtrsim$50 pc, indicating the potential for more discoveries with ongoing, dedicated searches.
 
\section{\emph{James Webb Space Telescope} Survey Simulations and Future Work}

The \emph{James Webb Space Telescope} will provide unprecedented imaging sensitivity. Observing 
at $\sim$4.5 $\mu$m with a coronagraph, the NIRCam instrument is expected 
to provide a contrast of 10$^{-5}$ at 1$^{\prime\prime}$ separations with a very low background limit.
To explore these capabilities in the context of M dwarfs, we have performed imaging survey simulations on 
our young M dwarf sample using the Monte Carlo methods described in \cite[Beichman et al.~(2010)]{Beichman2010}. Young planet
magnitudes were taken from extended COND03 models (\cite[Baraffe et al.~2003]{Baraffe2003}) covering 0.1 - 5 M$_{Jup}$. 
The simulated survey results reveal routine imaging sensitivity to planets $<$0.5 M$_{Jup}$ at $>$50 AU separations. In the best
cases, the simulations predict \emph{JWST}/NIRCam will detect $\sim$0.1 M$_{Jup}$ ($\sim$2 M$_{Nep}$) planets at $<$10 AU,
directly probing the known micro-lensing population.  

Our preliminary simulations indicate that 
\emph{JWST}/NIRCam imaging can routinely detect sub-Jovian mass planets around nearby, young M dwarfs. However, further work 
is warranted. New constraints on the M dwarf planet population are available and new planet evolution and atmosphere models 
will provide better predictions of young, low-mass planet luminosities. We plan to 
include these new results in our future simulations. Our continuing \emph{CASTOFFS} survey and improved survey
simulations will provide high priority targets for early \emph{JWST} GTO and GO proposals and pave the way for the first direct images
of low-mass gas-giant planets.


\begin{thebibliography}{}

\bibitem[Baraffe et al.~2003]{Baraffe2003}
{Baraffe, I., et al.} 2003,
\textit{A\&A}, 402, 701

\bibitem[Beichman et al.~(2010)]{Beichman2010}
{Beichman, C. A., et al.} 2010,
\textit{PASP}, 122, 162

\bibitem[Bowler et al.~2015]{Bowler2015}
{Bowler, B., et al.} 2015, 
\textit{ApJS}, 216, 7

\bibitem[Cassan et al.~2012]{Cassan2012}
{Cassan, A., et al.} 2012, 
\textit{Nature}, 481, 167

\bibitem[Kraus et al.~2014]{Kraus2014}
{Kraus, A.L., et al.} 2014, 
\textit{ApJ}, 147, 146

\bibitem[Schlieder et al.~2012]{Schlieder2012}
{Schlieder, J. E., et al.} 2012,
\textit{AJ}, 143, 80

\bibitem[Schlieder et al.~2015]{Schlieder2015}
{Schlieder, J. E., et al.} 2015,
\textit{Proceedings of CS18}, Editors: G. van Belle and H.C. Harris, 919

\bibitem[Torres et al.~2008]{Torres2008}
{Torres, C.A.O., et al.} 2008,
\textit{HSFR}, Editor: Bo Reipurth, 2, 757

\bibitem[Zuckerman \& Song 2004]{Zuckerman2004}
{Zuckerman, B. \& Song, I.} 2004, 
\textit{ARA\&A}, 42, 645






\end{thebibliography}
\end{document}